\documentclass[letter]{aa} % for the letters 
\usepackage{graphicx}
\usepackage{txfonts}
\usepackage{natbib}

\usepackage{hyperref}%[links=true, link={red}, cite={blue}]
\hypersetup{
    colorlinks = true,
    linkcolor = {red},
    citecolor = {blue}
}
\begin{document}

   \title{Asteroids' reflectance from $Gaia$ DR3:\\
   Artificial reddening at near-UV wavelengths}

   \author{F. Tinaut-Ruano
          \inst{1,2}
          \and
          E. Tatsumi\inst{1,2,3}
          \and
          P. Tanga\inst{4}
          \and
          J. de Le\'{o}n\inst{1,2}
          \and
          M. Delbo\inst{4}
          \and
          F. De Angeli\inst{5}
          \and
          D. Morate\inst{1,2}
          \and
          J. Licandro\inst{1,2}
          \and
          L. Galluccio\inst{4}
          }

   \institute{Instituto de Astrofísica de Canarias (IAC), C/ Vía Láctea, s/n, E-38205, La Laguna, Spain\\
   \email{fernando.tinaut@iac.es}
    \and
        Department of Astrophysics, University of La Laguna, Tenerife, Spain
     \and 
     Department of Earth and Planetary Science, University of Tokyo, 7-3-1 Hongo, Bunkyo-ku, 113-0033 Tokyo, Japan
     \and
     Universit\'{e} C\^{o}te d'Azur, Observatoire de la C\^{o}te d'Azur, CNRS, Laboratoire Lagrange, Bd de l'Observatoire, CS 34229, 06304 Nice Cedex 4, France 
     \and
     Institute of Astronomy, University of Cambridge, Madingley Road, Cambridge CB3 0HA, UK
             }

   \date{Received 04/10/2022; accepted 02/01/2023}

% \abstract{}{}{}{}{} 
% 5 {} token are mandatory
 
  \abstract
   {Observational and instrumental difficulties observing small bodies below 0.5 $\mu$m make this wavelength range poorly studied compared with the visible and near-infrared. 
   Furthermore, the suitability of many commonly used solar analogues, essential in the computation of asteroid reflectances, is usually assessed only in visible wavelengths, while some of these objects show spectra that are quite different from the spectrum of the Sun at wavelengths below 0.55 $\mu$m. Stars HD 28099 (Hyades 64) and HD 186427 (16 Cyg B) are two well-studied solar analogues that instead present spectra that are also very similar to the spectrum of the Sun in the wavelength region between 0.36 and 0.55 $\mu$m.}
   %Furthermore, the spectra of stars commonly used as solar analogues to compute asteroid reflectance spectra could be very similar to the spectrum of the Sun in visible wavelengths but some of them have a very different shape below 0.55 $\mu$m. Instead, Hyades 64 and 16 Cygnus B are two well studied solar analogues, properly characterized in near ultraviolet wavelengths in previous studies. Their spectra are very similar to that of the Sun over the complete visible and near-infrared spectral range, including the region between 0.36 and 0.55 $\mu$m.
   {We aim to assess the suitability in the near-ultraviolet (NUV) region of the solar analogues selected by the team responsible for the asteroid reflectance included in $Gaia$ Data Release 3 (DR3) and to suggest a correction (in the form of multiplicative factors) to be applied to the $Gaia$ DR3 asteroid reflectance spectra to account for the differences with respect to the solar analogue Hyades 64.}
   {To compute the multiplicative factors, we calculated the ratio between the solar analogues used by $Gaia$ DR3 and Hyades 64, and then we averaged and binned this ratio in the same way as the asteroid spectra in $Gaia$ DR3. We also compared both the original and corrected $Gaia$ asteroid spectra with observational data from the Eight Color Asteroid Survey (ECAS), one UV spectrum obtained with the Hubble Space Telescope (HST) and a set of blue-visible spectra obtained with the 3.6m Telescopio Nazionale Galileo (TNG). By means of this comparison, we quantified the goodness of the obtained correction.}
   {We find that the solar analogues selected for $Gaia$ DR3 to compute the reflectance spectra of the asteroids of this data release have a systematically redder spectral slope at wavelengths shorter than 0.55 $\mu$m than Hyades 64. We find that no correction is needed in the red photometer (RP, between 0.7 and 1 $\mu$m), but a correction should be applied at wavelengths below 0.55 $\mu$m, that is in the blue photometer (BP). After applying the correction, we find a better agreement between $Gaia$ DR3 spectra, ECAS, HST, and our set of  ground-based observations  with the TNG.}
   {Correcting the near-UV part of the asteroid reflectance spectra is very important for proper comparisons with laboratory spectra (minerals, meteorite samples, etc.) or to analyse quantitatively the UV absorption (which is particularly important to study hydration in primitive asteroids). The spectral behaviour at wavelengths below 0.5 $\mu$m of the selected solar analogues should be fully studied and taken into account for $Gaia$ DR4.}


   \maketitle
%
%-------------------------------------------------------------------

\section{Introduction}\label{sec:int}

Asteroid reflectance spectra and/or spectrophotometry provide(s) information on their surfaces' composition and the processes that modify their properties such as space weathering \citep{2015aste.book...43R}. Historically, the use of photoelectric detectors (or photometers), which are more sensitive at bluer wavelengths (e.g. < 0.5 $\mu$m), and the development of the standard $UBV$ photometric system \citep{1951ApJ...114..522J} led to the appearance of the first asteroid taxonomies in the 1970s \citep{1973BAAS....5..388Z,1975Icar...25..104C}, which contained information at blue-visible wavelengths or what we call near-UV (NUV). The introduction of the charge-coupled-devices (CCDs) in astronomy in the 1990s and later on contributed to the 'loss' of NUV information, as CCDs were much less sensitive at those wavelengths. Therefore, the large majority of the modern spectroscopic and spectrophotometric surveys cover the wavelength range from $\sim$0.5 $\mu$m up to 2.5 $\mu$m. Nevertheless, there are some exceptions. One of the first large surveys with information in the NUV is the Eight  Asteroid Survey (ECAS, \citealt{1985Icar...61..355Z}).  In this survey, we can find the photometry in eight broad-band filters between 0.34 to 1.04 $\mu$m for 589 minor planets, including two filters below 0.45 $\mu$m. These observations were used to develop a new taxonomy (see \citealt{1984PhDT.........3T}). Other recent catalogues, such as the Sloan Digital Sky Survey (SDSS) Moving Objects catalogue \citep{2002AJ....124.2943I}, the Moving Objects Observed from Javalambre (MOOJa) catalogue from the J-PLUS survey \citep{2021A&A...655A..47M}, or the Solar System objects observations from the SkyMapper Southern Survey \citep{2022A&A...658A.109S}, also include photometry in five, 12, and six filters between 0.3 and 1.1 $\mu$m with 104,449, 3122, and 205,515 objects observed, respectively. The new $Gaia$ data release 3 (DR3 hereafter) catalogue, which was released in June 2022, offers 60,518 objects binned in 16 wavelengths between 0.352 and 1.056 $\mu$m to mean reflectance
spectra.

Even though some laboratory measurements suggest the potential of the NUV absorption as a diagnostic region of hydrated and ferric material \citep{1979aste.book..688G, 1981PhDT........44F, 1985Icar...63..183F, 1996M&PS...31..321H, 2011Icar..212..180C, 2011Icar..216..309C, 2016M&PS...51..105H, 2021PolSc..2900723H}, a quantitative distribution of the NUV absorption among asteroids has not been discussed before \citep{2022arXiv220513917T}.  The small sensitivity of CCDs and the lower Sun's emission in NUV wavelengths make observations difficult. Moreover, the Rayleigh scattering by the atmosphere is stronger on shorter wavelengths, decreasing the signal-to-noise ratio (S/N) for the NUV region observed from the ground.
To compute the reflectance spectra, we needed to divide wavelength by wavelength of the measured spectra by the spectra of the Sun. As it is unpractical to observe the Sun with the same instrument used to observe asteroids, we used solar analogues (SAs), that is stars selected by their known similar spectra to that of the Sun. As the large majority of the spectroscopic and spectrophotometric surveys cover the wavelength range that goes from the visible to near-infrared (NIR), the most commonly used SAs are well characterised at those wavelengths but they can behave very differently in the NUV. This flux difference at bluer wavelengths can introduce systematic errors in the asteroid reflectance spectra. A good example is the work by \cite{2016Icar..266...57D}, where they searched for the presence of F-type asteroids in the Polana collisional family since the parent body of the family, asteroid (142) Polana, was classified as an F type. The authors obtained reflectance spectra in the NUV of the members of the family, finding that the large majority were classified as B types. As most of the observers, they used SAs that were widely used by the community. Interestingly, after obtaining the asteroid reflectances again using only Hyades 64 as the SA, \cite{2022arXiv220513917T} found that the large majority of the observed members of the Polana family were indeed F types and not B types. This evidences the importance of using adequate SAs when observing in the NUV, and it has been the main motivation for this work.

In this Letter, we present a comparison between the SAs selected to compute the reflectance spectra in the frame of the data processing of $Gaia$ DR3 \citep{2022arXiv220612174G} and Hyades 64. We analyse the results from this comparison and propose a multiplicative correction that can be applied to the archived asteroids' reflectance spectra. We finally tested it by comparing corrected $Gaia$ reflectance spectra with ground-based observations that have also been corrected against the same SA (ECAS survey, TNG spectra) and with one observation with the Hubble Space Telescope (HST).
%--------------------------------------------------------------------
\section{Sample}\label{sec:sam}
\subsection{Solar analogues in $Gaia$ DR3}
The $Gaia$ DR3 catalogue \citep{2022arXiv220612174G} gives access to internally and externally calibrated mean spectra for a large subset of sources. Internally calibrated spectra refer to an internal reference system that is homogeneous across all different instrumental configurations, while externally calibrated spectra are given in an absolute wavelength and flux scale \cite[see][for more details]{2022arXiv220606143D, 2022arXiv220606205M}. Epoch spectra (spectra derived from a single observation rather than averaging many observations of the same source) are not included in this release. 
%In this paper, we will make use of externally calibrated mean spectra when comparing the $Gaia$ DR3 spectra of SAs with the spectrum of the Sun downloaded from SORCE\footnote{\url{http://lasp.ado.edu/data/sorce/ssi_data/composite/sorce_ssi_latest.txt.zip}}.
For this Letter, we relied on internally calibrated data when computing the correction for the $Gaia$ reflectances to ensure consistency and to avoid artefacts that could appear when dividing two externally calibrated spectra, as they are polynomial fits.
%For the $Gaia$ DR3 catalogue, \citet{2022arXiv220612174G} generated internally calibrated spectra, averaged along epochs, in both available photometers, the Blue Photometer (BP) and the Red Photometer (RP), that converts raw pixel data into an internal system homogeneous across all different instruments set-ups \citep[see more in][]{2022arXiv220612174G}. After this internal calibration, a polynomial fitting transforms pseudo-wavelengths to wavelengths, this final product named externally calibrated data. They also provide the mean spectrum of a group of SAs, \cite[information about the internal and external calibrations in][]{2022arXiv220606143D, 2022arXiv220606205M}.

To select the SAs, the $Gaia$ team did a bibliographic search and selected a list of stars that are widely used as solar analogues for asteroid spectroscopy \citep{2002Icar..158..106B,2004Icar..172..179L,2004A&A...418.1089S,2007Icar..190..622F, 2014A&A...572A.106P,2018P&SS..157...82P, 2019A&A...627A.124P,2019Icar..322..227L}. First of all, we note that the star identified as 16 Cygnus B in \cite{2022arXiv220612174G} is in fact 16 Cygnus A and that the parameters in their Table C.1. correspond to those of 16 Cygnus A. Luckily enough, the spectrum of 16 Cygnus B was also available in DR3. Among the referenced works, only \cite{2004A&A...418.1089S} carried out a search for SAs by comparing their spectra to that of the Sun down to 0.385 $\mu$m. The rest simply used G2V stars or cited previous works that presented SAs, as in \cite{ 1978A&A....63..383H}. In this later work, Hardorp selected SAs by comparing their spectra with the spectrum of the Sun using wavelengths down to 0.36 $\mu$m. He  highlighted the variations that can exist at NUV wavelengths even between stars of the same spectral class. 

\subsection{Asteroids in $Gaia$ DR3}
Among the $Gaia$ DR3 products for Solar System objects (SSOs), neither the internally nor the externally calibrated spectra are available to the community, as is the case for the stars. This is due to a specific choice of the Data Processing and Analysis Consortium (DPAC) caused by the difficulty of calculating those quantities owing to the intrinsic variability and proper motion of SSOs. Instead, for each SSO and each epoch, the nominal, pre-launch dispersion function was used to convert pseudo-wavelengths to physical wavelengths. The reflectance spectra were calculated by dividing each epoch spectrum by the mean of the SAs selected and then averaging over the set of epochs. After that, a set of fixed wavelengths every 44 nm in the range between 374 and 1034 nm was defined, with a set of bins centred at those wavelengths and with a size of 44 nm. For each bin (a total of 16 are provided), a $\sigma$-clipping filter was applied and a weighted average using the inverse of the standard deviation as weight was obtained. Finally, the reflectances were normalised to unity using the value at 550$\pm$25 nm. This final product is the only one available in DR3.

\subsection{Hyades 64 \& 16 Cyg B}
As mentioned in Sect. 2.1, \cite{1978A&A....63..383H} concluded that Hyades 64 and 16 Cyg B are two of the four stars that exhibit 'almost indistinguishable' NUV spectra (quoting the author's words) from the spectrum of the Sun. This was confirmed in subsequent papers from the same author \citep{1980A&A....88..334H,1980A&A....91..221H} and from other researchers \citep{1996A&ARv...7..243C,1997ApJ...482L..89P,Farnham2000,2004A&A...418.1089S}. We used these two stars as a 'reference' to compute the correction factor to be applied to the $Gaia$ DR3 asteroids spectra, as they are in the list of SAs selected by \cite{2022arXiv220612174G}. The methodology is described in the following section. We note that the obtained correction factor using Hyades 64 as opposed to 16 Cygnus B differs less than 0.5\%. We, therefore, decided to use Hyades 64, as it was the star that was used for both the ECAS survey and our ground-based observations. 

%TODO: buscar los otros autores que recomiendan Hyades 64 y añadir aqui.

%--------------------------------------------------------------------
%\begin{figure}[!ht]
%\centering
%\includegraphics[width=0.95\columnwidth]{GAIA_SA_sun_diff.pdf}
%\caption{Pearson coefficient, between binned reflectance spectra of solar analogues and the Sun. Errors bars are computed as the standard deviation of 1000 Pearson coefficient calculated over randomly selected 85\% of the spectral points. Blue stars are SAs proposed by the $Gaia$ team in \cite{2022arXiv220612174G} and black stars are the two favourites stars selected by \cite{1978A&A....63..383H}. $^1$Note that the star identified as 16 Cygnus B in \cite{2022arXiv220612174G} is in fact 16 Cyg A. See text for more details.}
%\label{f:Met:diff_GAIA_SA_Hyd_64_sun}
%\end{figure}

\section{Methodology}\label{sec:met}

\begin{figure*}[!h]
\centering
\includegraphics[width=0.98\textwidth]{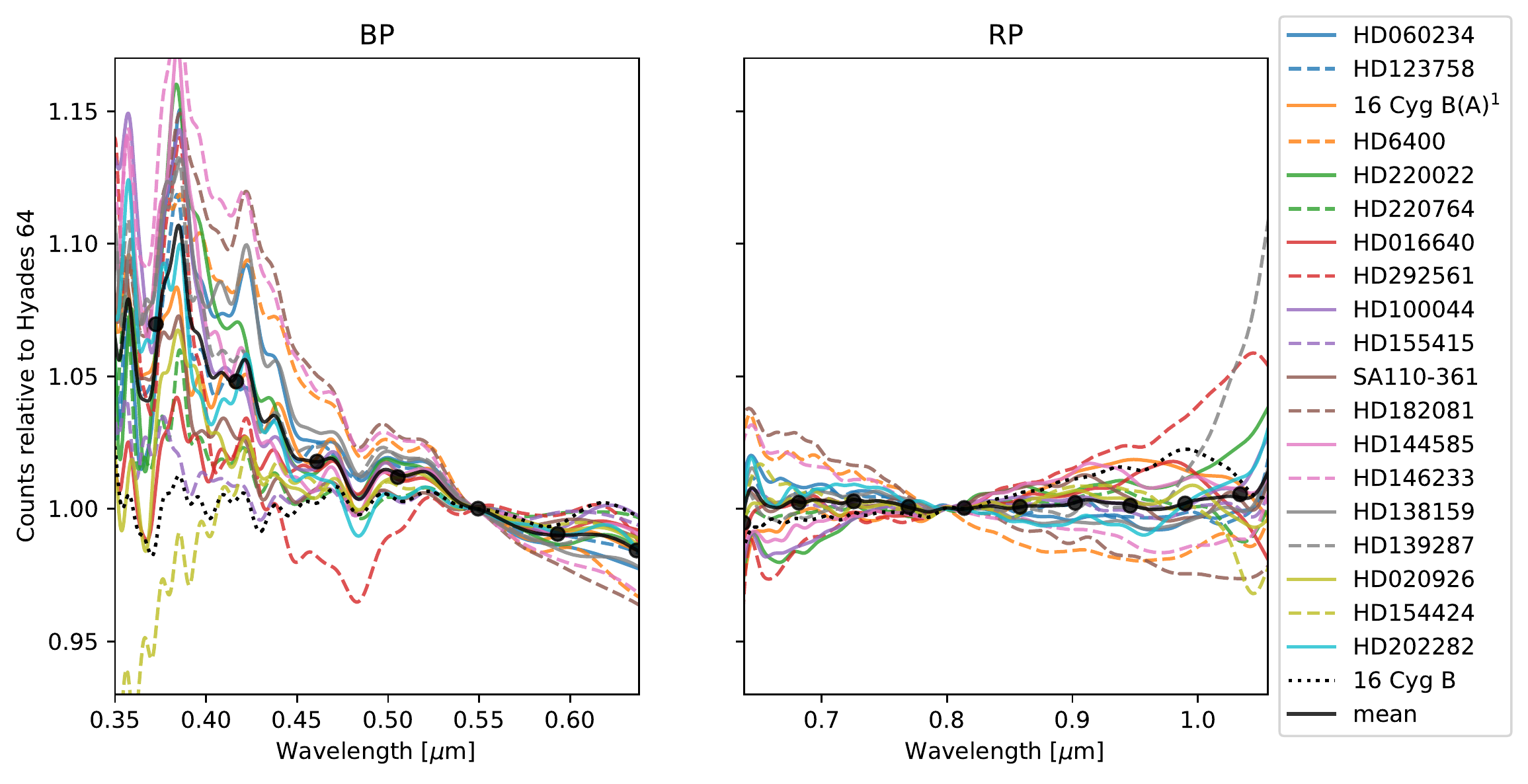}
\caption{Ratio between the internally calibrated spectra of each of the $Gaia$ SAs and Hyades 64 in the blue photometer (BP, left panel) and the red photometer (RP, right panel). We also plotted the ratio of the mean $Gaia$ SA and Hyades 64 (black solid line) and the binned version of this ratio at the wavelengths provided for SSO in $Gaia$ DR3 (black dots).\protect\\ $^1$ We note that the star identified as 16
Cygnus B in Gaia Collaboration et al. (2022) is in fact 16 Cyg A (see the main
text for more details).}
\label{f:Met:GaiaSA_ourSA_bp_rp}
\end{figure*}

%\subsection{$Gaia$'s SAs in the NUV: externally calibrated data}
%To study the behaviour of the SAs selected by the DPAC at blue wavelengths, we normalized  their externally calibrated spectra at 0.55 $\mu$m, and we binned the resulting spectra at the same wavelengths as done for the asteroid reflectance spectra. We proceeded in the same way with the Sun spectrum downloaded from SORCE$^1$. Finally, we compute the Pearson coefficient for every (star, sun) pair. This coefficient is independent of the magnitude of the stars as takes into account only the linear relationship between the SA and the Sun.

%To estimate errors of the Pearson coefficients we compute it over 85\% randomly selected points in the SA spectra 1000 times. The mean and the standard deviation are plotted for each SA in Fig. \ref{f:Met:diff_GAIA_SA_Hyd_64_sun}. We note that the star identified as 16 Cygnus B in \cite{2022arXiv220612174G} is in fact 16 Cygnus A and that the parameters in their Table C.1. correspond to those of 16 Cygnus A. Luckily enough, the spectrum of 16 Cygnus B was also available in the DR3. In agreement with Hardorp results, Hyades 64 and 16 Cygnus B (in black in Fig. \ref{f:Met:diff_GAIA_SA_Hyd_64_sun}) are among the stars that are more similar to the Sun. To be able to compare our correction with \cite{2022arXiv220513917T}, we will hereafter use Hyades 64 as the reference SA. The other star selected by Hardorp, 16 Cyg B, is too bright to obtain proper spectra from ground with big telescopes. 

\begin{figure}
\centering
\includegraphics[width=\columnwidth]{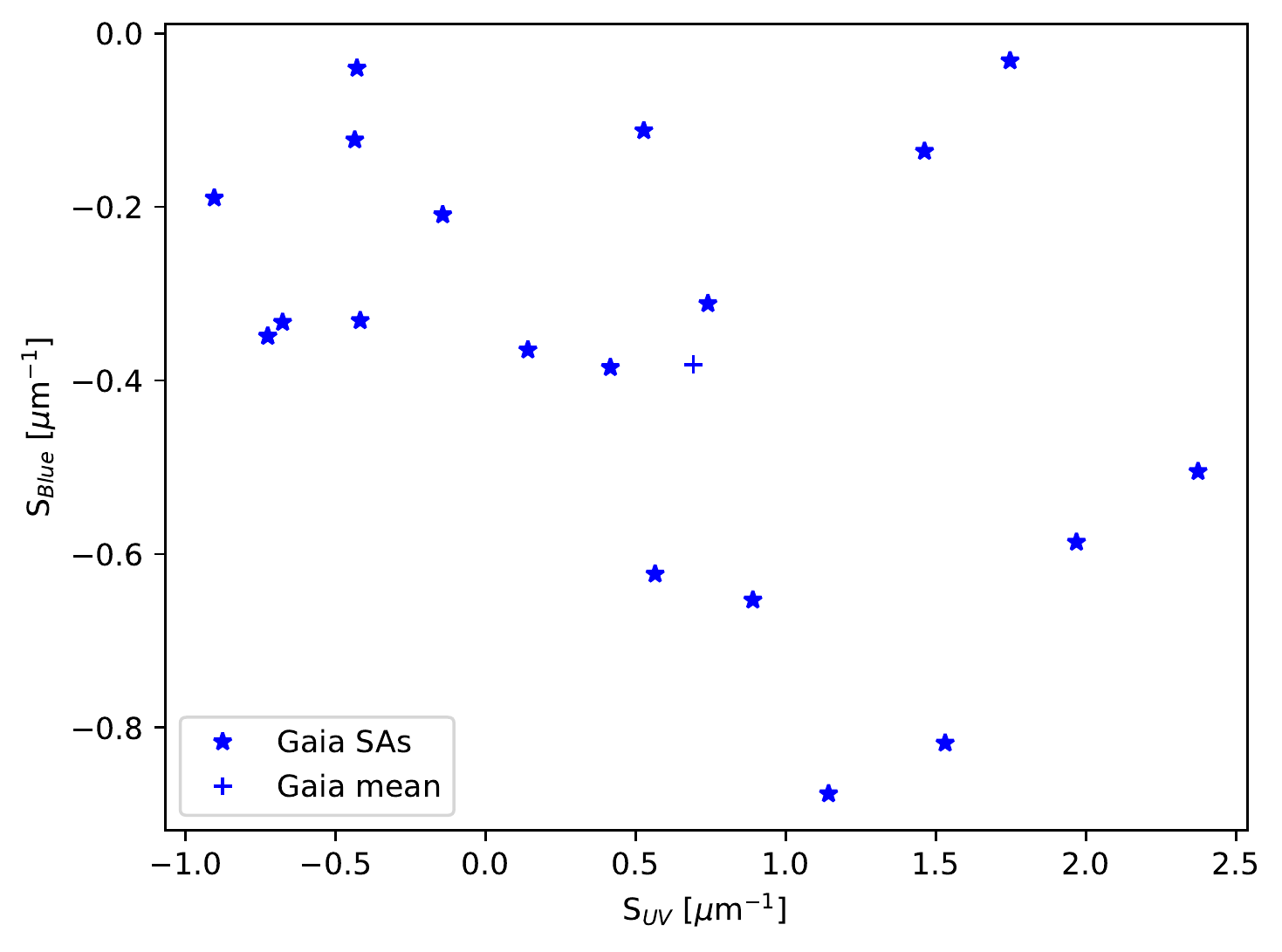}
\caption{Slopes introduced by each of the SAs in the $Gaia$ sample (blue stars) and their mean (blue cross), compared to Hyades 64.  We note that $S_{\textnormal{Blue}}$ was computed in the 0.4--0.55 $\mu$m range, while $S_{\textnormal{UV}}$ was computed using wavelengths below 0.4 $\mu$m.}
\label{f:Met:GaiaSA_ourSA_slope}
\end{figure}

\subsection{Computing the correction factor: Internally calibrated data}
In order to compute a correction applicable to the $Gaia$ DR3 reflectances, we proceeded as follows: first, using the internally calibrated data, we computed the ratio between the $Gaia$ sample of SAs, as well as the mean spectrum of these SAs, and Hyades 64 (Fig. \ref{f:Met:GaiaSA_ourSA_bp_rp}). As we can observe in the right panel of Fig. \ref{f:Met:GaiaSA_ourSA_bp_rp}, which corresponds to the red photometer (RP), the deviation from the unity of the ratio between $Gaia$'s mean SA and Hyades 64 (black line) is always below 1\%. Therefore, this mean spectrum can confidently be used to obtain the reflectance spectra of asteroids above 0.55 $\mu$m.

However, the situation in the the blue photometer (BP) is quite different. We can see in the left panel of Fig. \ref{f:Met:GaiaSA_ourSA_bp_rp} that the deviation from the unity of the above defined ratio can reach values of up to 10\%, indicating that the mean spectrum of the SAs used in $Gaia$ DR3 differs significantly from Hyades 64 at wavelengths below 0.55 $\mu$m. The biggest effect when using this mean spectrum to obtain asteroids' reflectance spectra is the introduction of a systematic (and not real) positive slope, in particular in the range between 0.4 and 0.55 $\mu$m, mimicking a drop in reflectance below 0.55 $\mu$m. Furthermore, the division by this mean spectrum can also introduce a 'fake' absorption around 0.38 $\mu$m. We have quantified this spectral slope in two separate wavelength ranges, trying to reproduce the observed behaviour of the ratio: one slope between 0.4 to 0.55 $\mu$m, which we named $S_{\textnormal{Blue}}$, and another one for wavelengths below 0.4 $\mu$m, named $\mu$m $S_{\textnormal{UV}}$. The obtained values for the individual SAs used in $Gaia$ DR3 (blue stars), as well as for the mean spectrum (blue cross) are shown in Fig. \ref{f:Met:GaiaSA_ourSA_slope}. For the mean spectrum of the SAs used in $Gaia$ DR3, we found that the introduced slopes are $S_{\textnormal{Blue}}$ = -0.38 $\mu$m$^{-1}$ and $S_{\textnormal{UV}}$ = 0.69 $\mu$m$^{-1}$.

From this analysis, we conclude that a correction is needed in the NUV wavelengths, that is below 0.55 $\mu$m. To arrive at the multiplicative correction factors, we binned the ratio between the mean spectra of SAs selected by the DPAC and Hyades 64, using the same wavelengths and bin size as the ones adopted for the asteroid reflectance spectra in the $Gaia$ DR3 (see Sect. \ref{sec:sam}). In this way, the users can easily correct the asteroid spectra at NUV wavelengths. The obtained values are shown in Table \ref{latablita}. 

\begin{table}
\caption{Multiplicative correction factors for $Gaia$ asteroid binned spectra. We include the wavelengths below 0.55 $\mu$m.}
\label{latablita}
\centering
\begin{tabular}{c c}   
\hline\hline
Wavelength ($\mu$m) & Correction factor\\
\hline
0.374 & 1.07\\
0.418 & 1.05\\
0.462 & 1.02\\
0.506 & 1.01\\
0.550 & 1.00\\
%0.594 & 0.99
\hline
\end{tabular}
\end{table}

\subsection{Comparison of corrected reflectances with existing data}
To correct the artificial slopes introduced by the use of the mean $Gaia$ SAs, we multiplied the binned asteroid reflectance spectra below 0.55 $\mu$m by the corresponding correction factors. We compared the corrected $Gaia$ spectra with spectra or spectrophotometry of the same asteroids obtained using other facilities. As a first step, we selected only those $Gaia$ asteroid spectra with a S/N $>$ 160, as we detected a systematic decrease in spectral slope values at blue wavelengths with decreasing S/N for objects with a smaller S/N than 150. We then selected spectrophotometric data from the ECAS survey for asteroids that have more than one observation, and NUV spectra obtained with the Telescopio Nazionale Galileo (TNG) and previously published by \cite{2022arXiv220513917T}. The resulting comparison dataset is shown in Fig. \ref{f:Res:asteroids_corr}, where the red lines correspond to the original $Gaia$ reflectances, black lines are the corrected ones, dark blue lines correspond to ECAS data, and TNG spectra are shown in light blue.
%For the sake of coherency, and to avoid potential differences due to problems associated with ground-based observations, we selected only those objects presenting a good agreement between ECAS and TNG measurements. 
As can be seen, the corrected reflectances are in better agreement with the ECAS and TNG data than the original ones. We also included the UV spectrum of asteroid (624) Hector downloaded from the ESA archive using the python package astroquery.esa.hubble\footnote{\url{https://astroquery.readthedocs.io/en/latest/esa/hubble/hubble.html}}. It was obtained with STIS at HST \citep{2019AJ....157..161W}. We converted the flux to reflectance using the spectrum of the Sun provided for the STIS instrument\footnote{\url{https://archive.stsci.edu/hlsps/reference-atlases/cdbs/current_calspec/sun_reference_stis_002.fits}}. We note that even after the correction, some asteroids show discrepancies with the reference data. This is discussed in the next section.

%--------------------------------------------------------------------
\section{Results and discussion}\label{sec:res}
\begin{figure}
\centering
\includegraphics[width=\columnwidth]{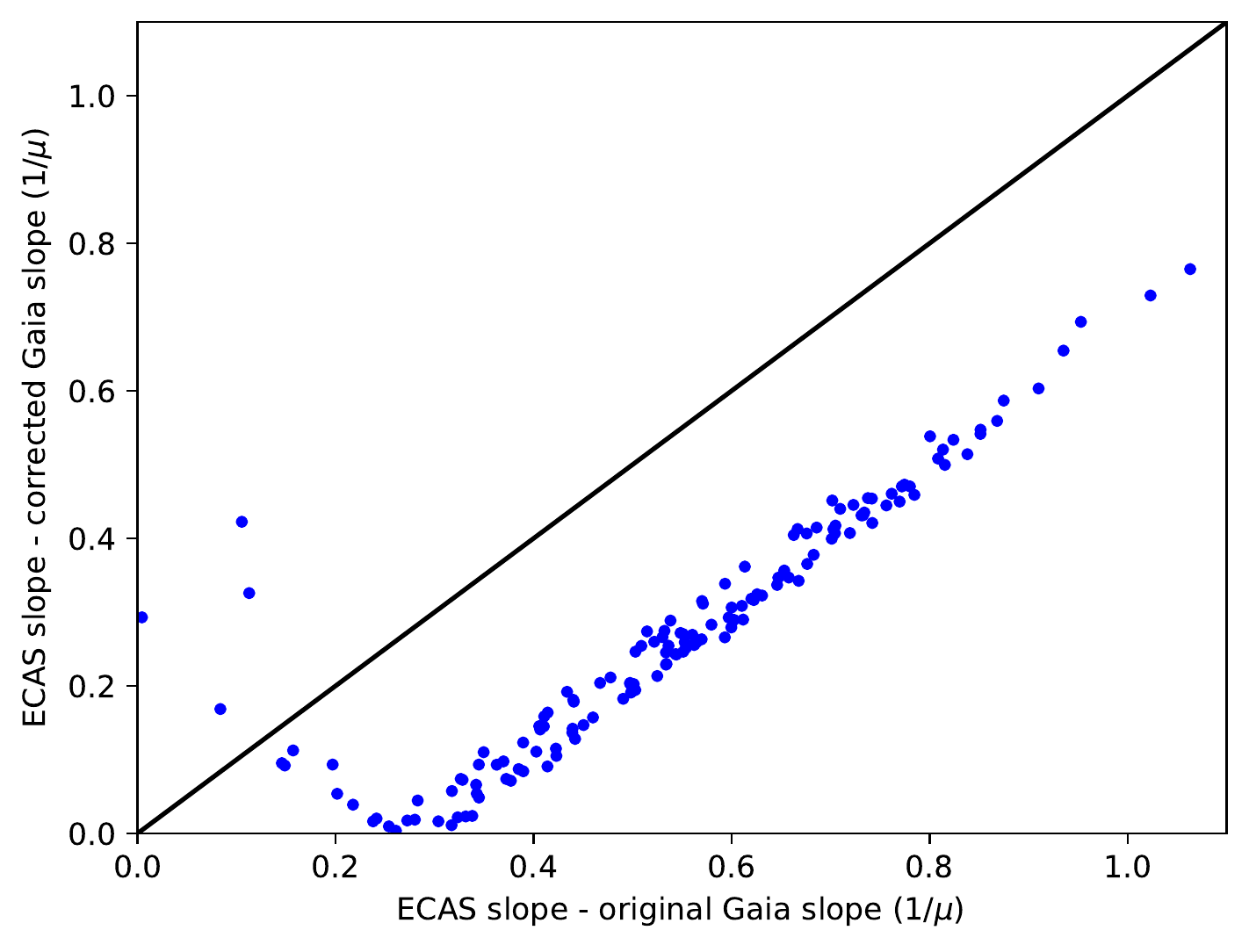}
\caption{Difference between the blue slope for ECAS and for $Gaia$ original data (x-axis) and corrected data (y-axis) in the comparison sample.}
\label{f:Res:corr_goddness}
\end{figure}

We have shown that the artificial slope introduced at blue wavelengths in the $Gaia$ DR3 asteroid data due to the selected SAs is -0.38 $\mu$m$^{-1}$ in the range between 0.4 to 0.55 $\mu$m and 0.69 $\mu$m$^{-1}$ below 0.4 $\mu$m. Following \cite{1985Icar...61..355Z}, the $b$ and $v$ filters of the ECAS survey have central effective wavelengths of 0.437 and 0.550 $\mu$m, respectively. According to \cite{1984PhDT.........3T}, the ($b$-$v$) colours of the mean F and B taxonomical classes are -0.049 and -0.015 magnitudes, respectively. Transforming these colours to relative reflectances results in 1.046 and 1.014, which gives slopes of -0.407 and -0.124 $\mu$m$^{-1}$ between 0.437 and 0.55 $\mu$m. Therefore, the difference between these computed slopes for F and B types (-0.283 $\mu$m$^{-1}$) is smaller than the artificial slope introduced by the use of the mean SA of $Gaia$, implying that unless we apply the correction proposed in this Letter, asteroids can be easily misclassified as B types when actually being F types (see the described example in the Introduction for the case of members of the Polana family).\\
To test and quantify the goodness of our proposed correction, we computed the spectral slope between 0.437 and 0.55 $\mu$m for the ECAS comparison dataset, and between 0.418 and 0.55 for $Gaia$ original and corrected spectra. In Fig. \ref{f:Res:corr_goddness} we plotted the difference between those slopes. After applying our correction factor, we could see that the large majority (148 out of 152) of the asteroids have more similar slopes to those of ECAS.\\
Nevertheless, our correction has limitations. First, we were testing its goodness over space-based observations using ground-based observations. For wavelengths down to 0.3 $\mu$m, ground-based observations present some difficulties, mainly due to the atmospheric absorption and the lower sensitivity of the detectors. Furthermore, $Gaia$ observations at those wavelengths also have other artifacts that we do not fully understand, such as the detected strong decrease in the spectral slope below S/N ~150. Another point to consider when comparing asteroid spectra observed in different epochs is the effect of the different viewing geometries. This difference in the viewing geometry, and thus, in the phase angle, causes a change in the spectral slope known as phase reddening or phase coloring  \cite{2022A&A...667A..81A}. This effect has not been well studied at blue wavelengths. Still, even in the event that we were able to correct it, $Gaia$'s spectra are, on average, over different epochs and the information on the phase angle values is not provided.\\

%-----------------------------------------------------------------
\section{Conclusions}\label{sec:con}

We have found that the use of the SAs selected to compute the reflectance spectra of the asteroids in $Gaia$ DR3 introduces an artificial reddening in the spectral slope below 0.5 $\mu$m, that is an artificial drop in reflectance. By comparing those SAs with Hyades 64, one of the best characterised SAs at NUV wavelengths, we obtain multiplicative correction factors for each of the reflectance wavelengths below 0.55 $\mu$m (a total of four) that can be applied to the asteroids' reflectance spectra in $Gaia$ DR3. By applying this correction, we found a better agreement between the $Gaia$ spectra and other data sources such as ECAS. The behaviour of the SAs in the red wavelengths is in agreement with Hyades 64 within 1\%. This was somehow expected, as the majority of the SAs used by the $Gaia$ team were previously tested and widely used by the community to obtain visible reflectance spectra of asteroids, typically beyond 0.45--0.5 $\mu$m. 

Correcting the NUV part of the asteroid reflectance spectra is fundamental to study the presence of the UV absorption, which has been associated with hydration in primitive asteroids, or to discriminate between B and F types, which are two taxonomical classes that have proven to have very distinct polarimetric properties. The NUV region has not yet been fully exploited for asteroids and, in this way, $Gaia$ spectra constitute a major step forward in our understanding of these wavelengths.

\begin{acknowledgements}
FTR, JdL, ET, DM, and JL acknowledge support from the Agencia Estatal de Investigaci\'{o}n del Ministerio de Ciencia e Innovaci\'{o}n (AEI-MCINN) under the grant 'Hydrated Minerals and Organic Compounds in Primitive Asteroids' with reference PID2020-120464GB-100.\\
FTR also acknowledges the support from the COST Action and the ESA Archival Visitor Programme.\\
DM acknowledges support from the ESA P3NEOI programme (AO/1-9591/18/D/MRP).\\
This work has made use of data from the European Space Agency (ESA) mission
{\it Gaia} (\url{https://www.cosmos.esa.int/gaia}), processed by the {\it Gaia}
Data Processing and Analysis Consortium (DPAC,
\url{https://www.cosmos.esa.int/web/gaia/dpac/consortium}). Funding for the DPAC
has been provided by national institutions, in particular, the institutions
participating in the {\it Gaia} Multilateral Agreement.\\
The work of MD is supported by the CNES and by the project Origins of the French National Research Agency (ANR-18-CE31-0014).\\
F. De Angeli is supported by the United Kingdom Space Agency (UKSA) through the grants ST/X00158X/1 and ST/W002469/1.

\end{acknowledgements}

\bibliographystyle{aa}
\bibliography{main}

\appendix
\section{Comparison figures}\label{appendix}
In this appendix, we are showing the spectra of asteroids that have at least two observations from the ground from ECAS (dark blue) or from TNG (light blue) and also have available spectra in Gaia DR3. We plotted the original (red) and the corrected (black) version of the Gaia spectra together. For asteroid 624 (Hector), we also added an observation from HST (further information can be found in the main text).
\begin{figure*}
\centering
\includegraphics[width=0.95\textwidth]{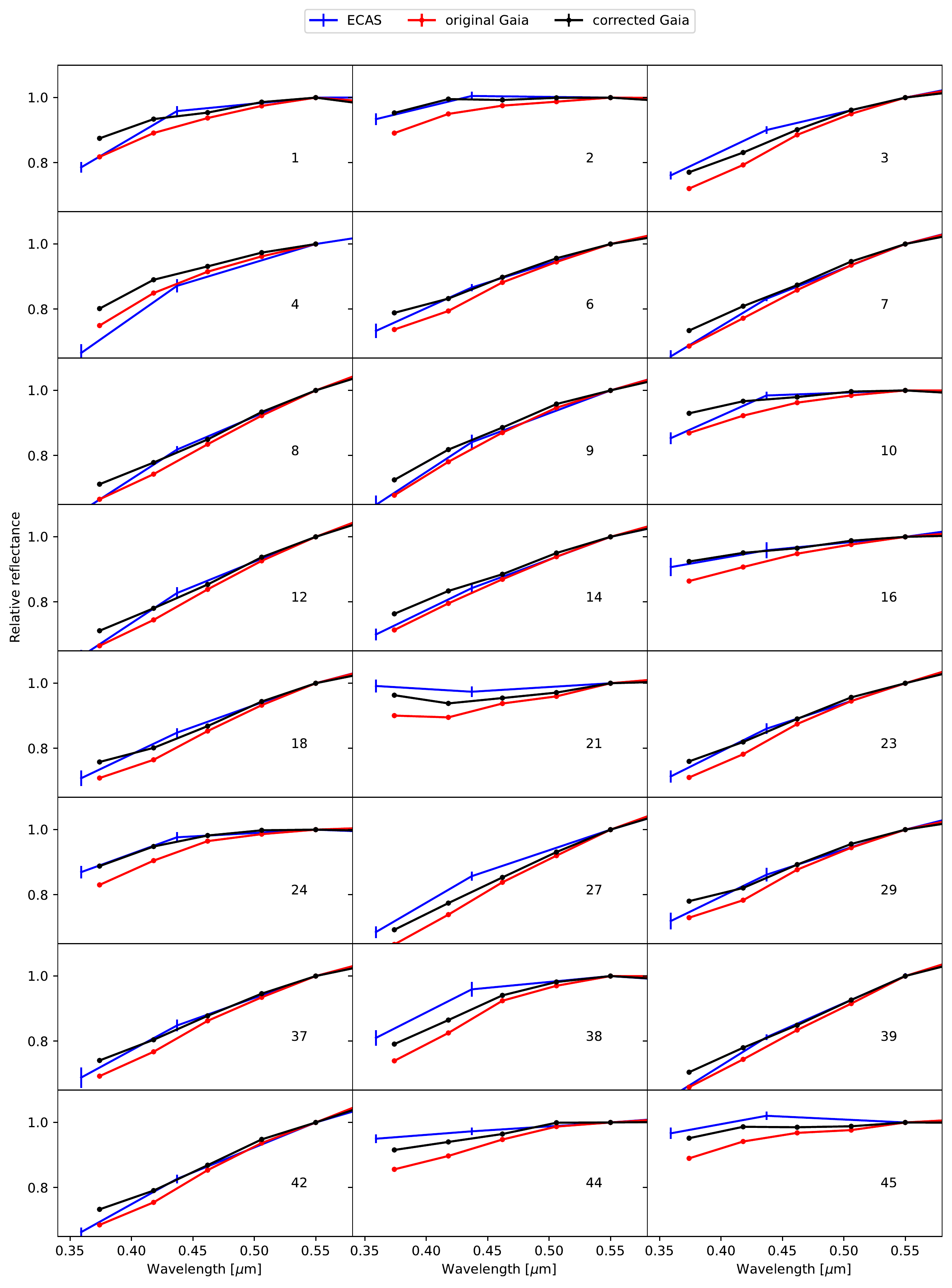}
\caption{Comparison between ground-based observations from the Eight  Asteroid Survey (ECAS, dark blue line), TNG observations (light blue line) original $Gaia$ data (red line), and corrected data (black line). We also included a UV spectrum of asteroid (624) downloaded from ESA archive and obtained with the instrument STIS, on board the Hubble Space Telescope (HST).}
\label{f:Res:asteroids_corr}
\end{figure*}

\clearpage

\begin{center}
\centering
\includegraphics[width=0.95\textwidth]{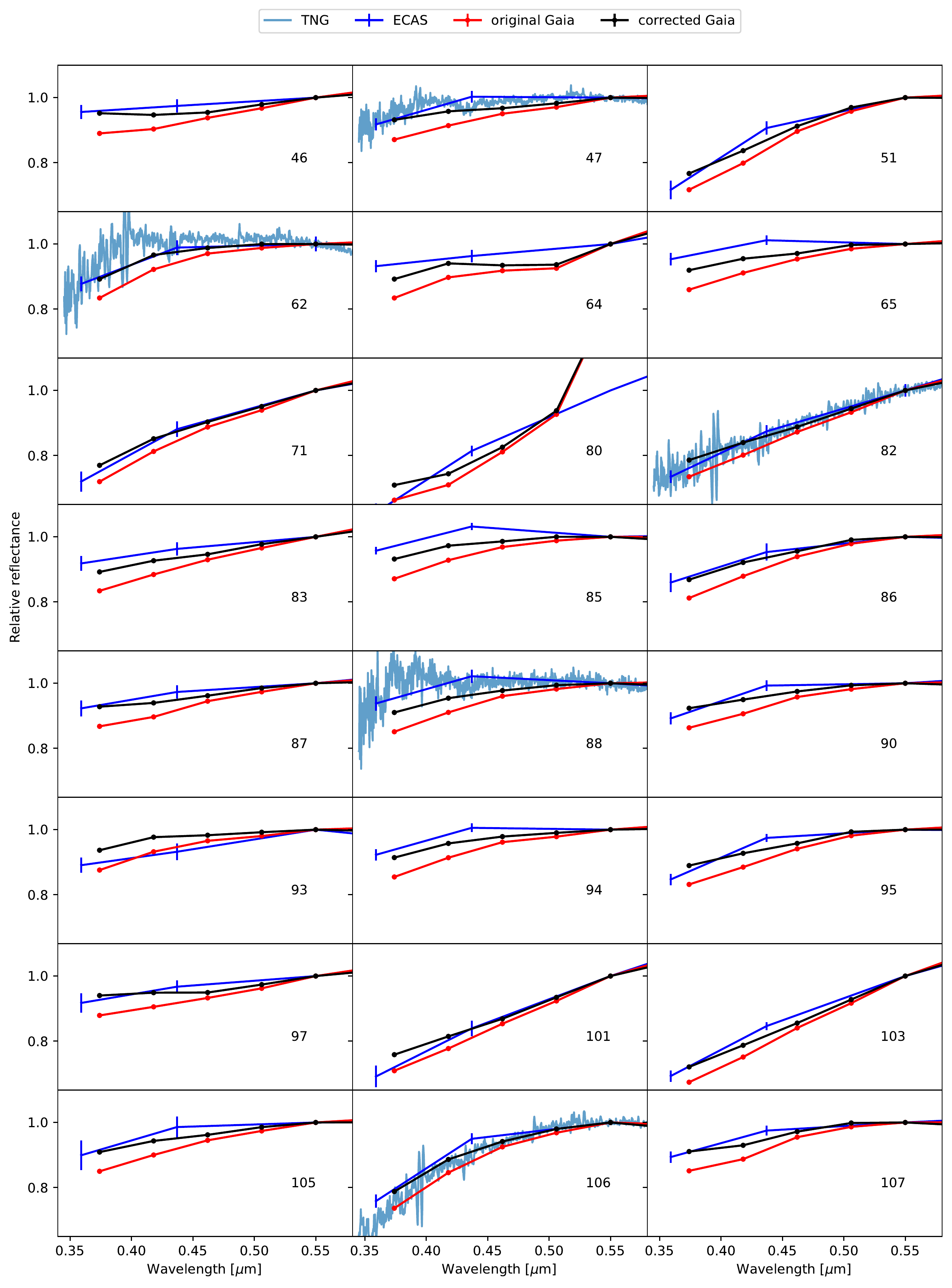}
\end{center}

\clearpage

\begin{center}
\centering
\includegraphics[width=0.95\textwidth]{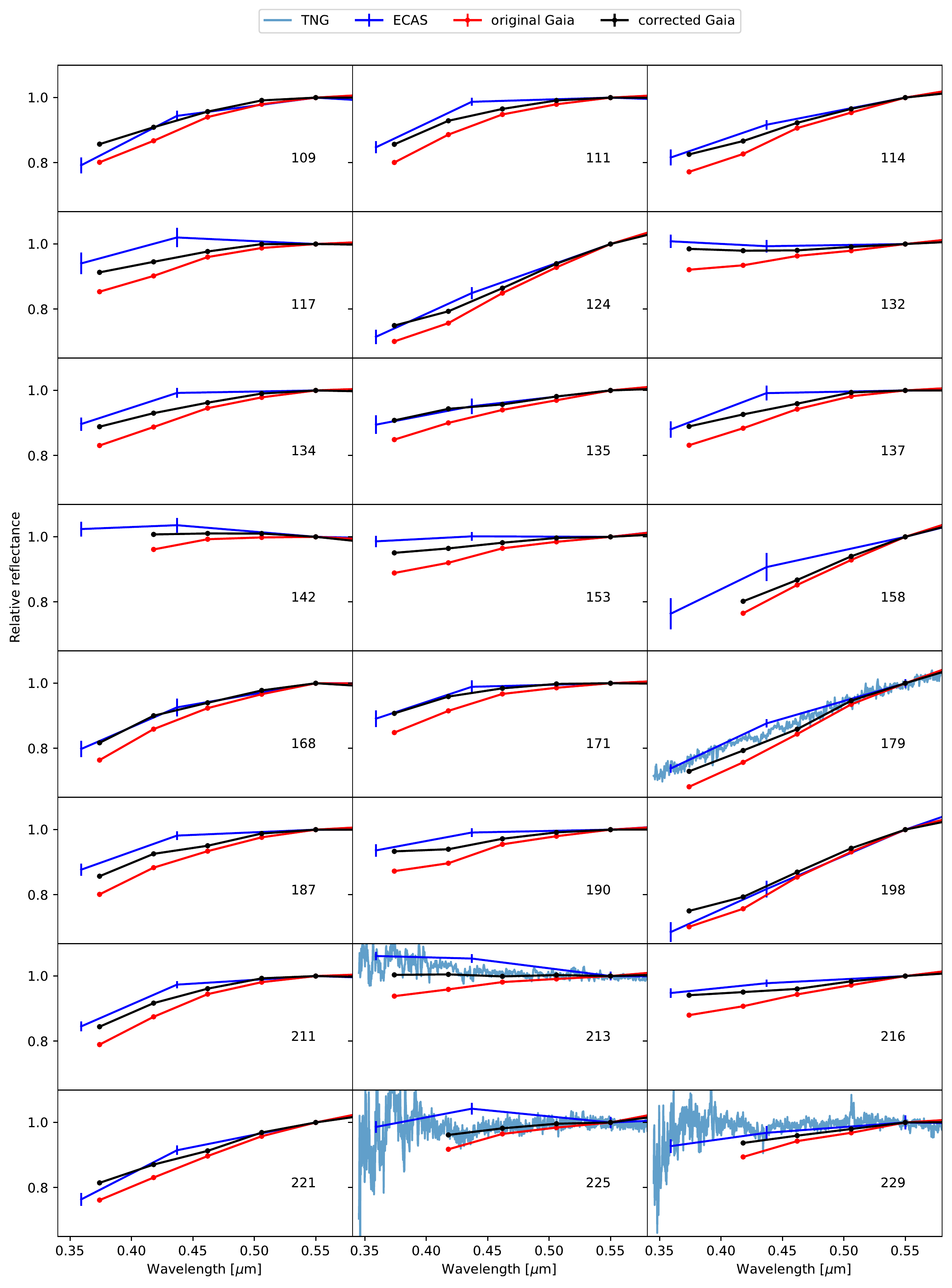}
\end{center}

\clearpage

\begin{center}
\centering
\includegraphics[width=0.95\textwidth]{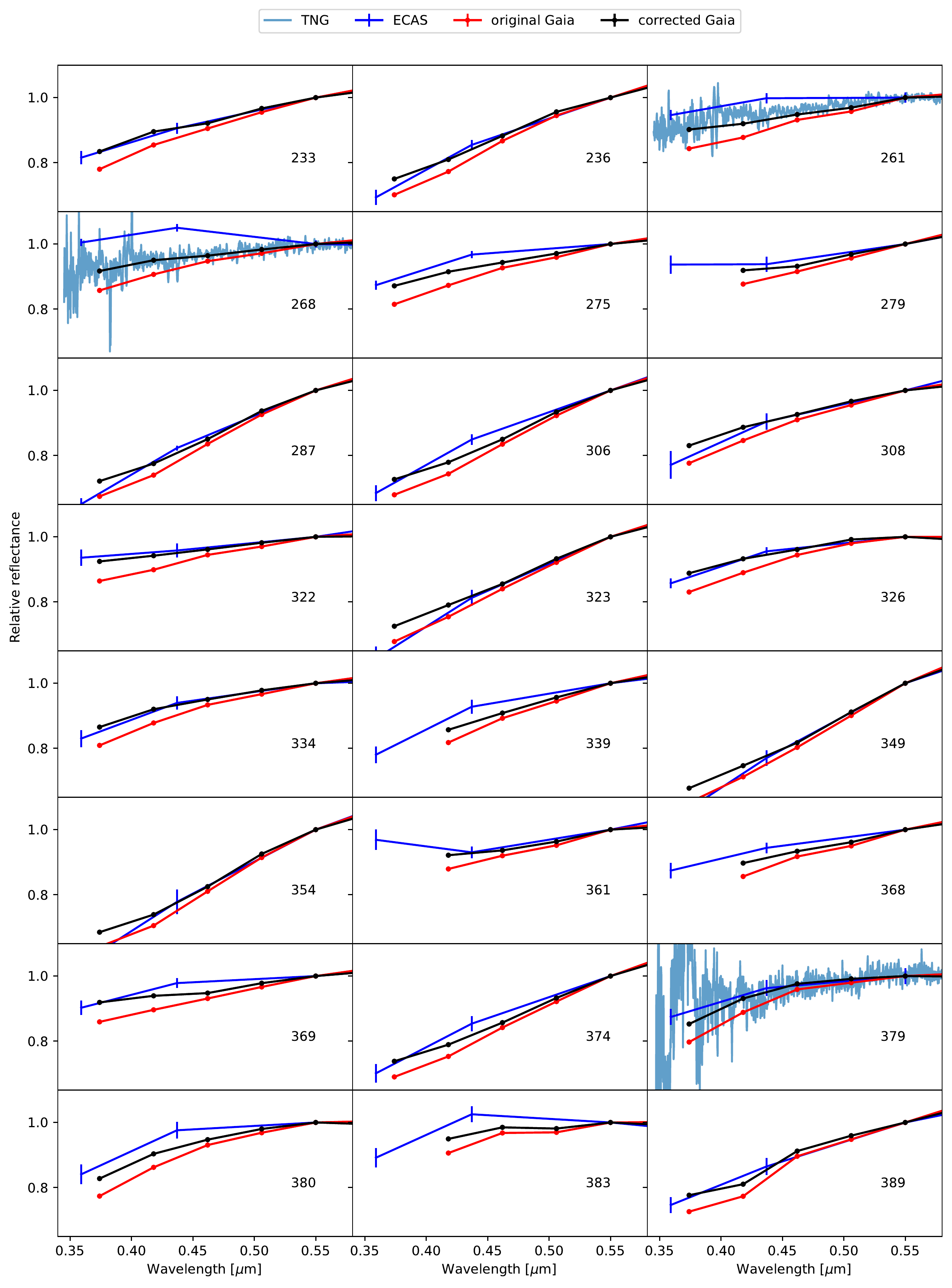}
\end{center}

\clearpage

\begin{center}
\centering
\includegraphics[width=0.95\textwidth]{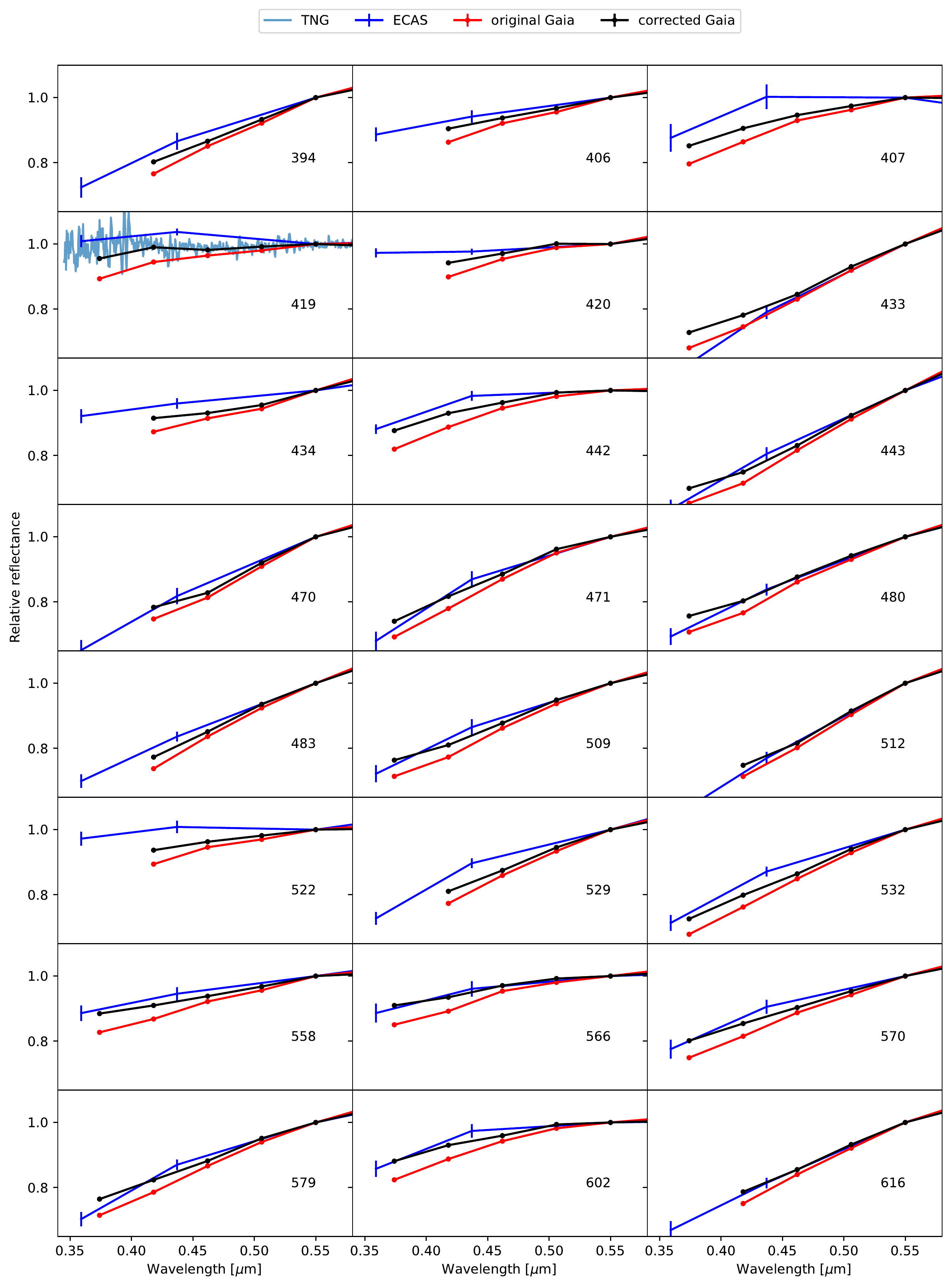}
\end{center}

\clearpage

\begin{center}
\centering
\includegraphics[width=0.95\textwidth]{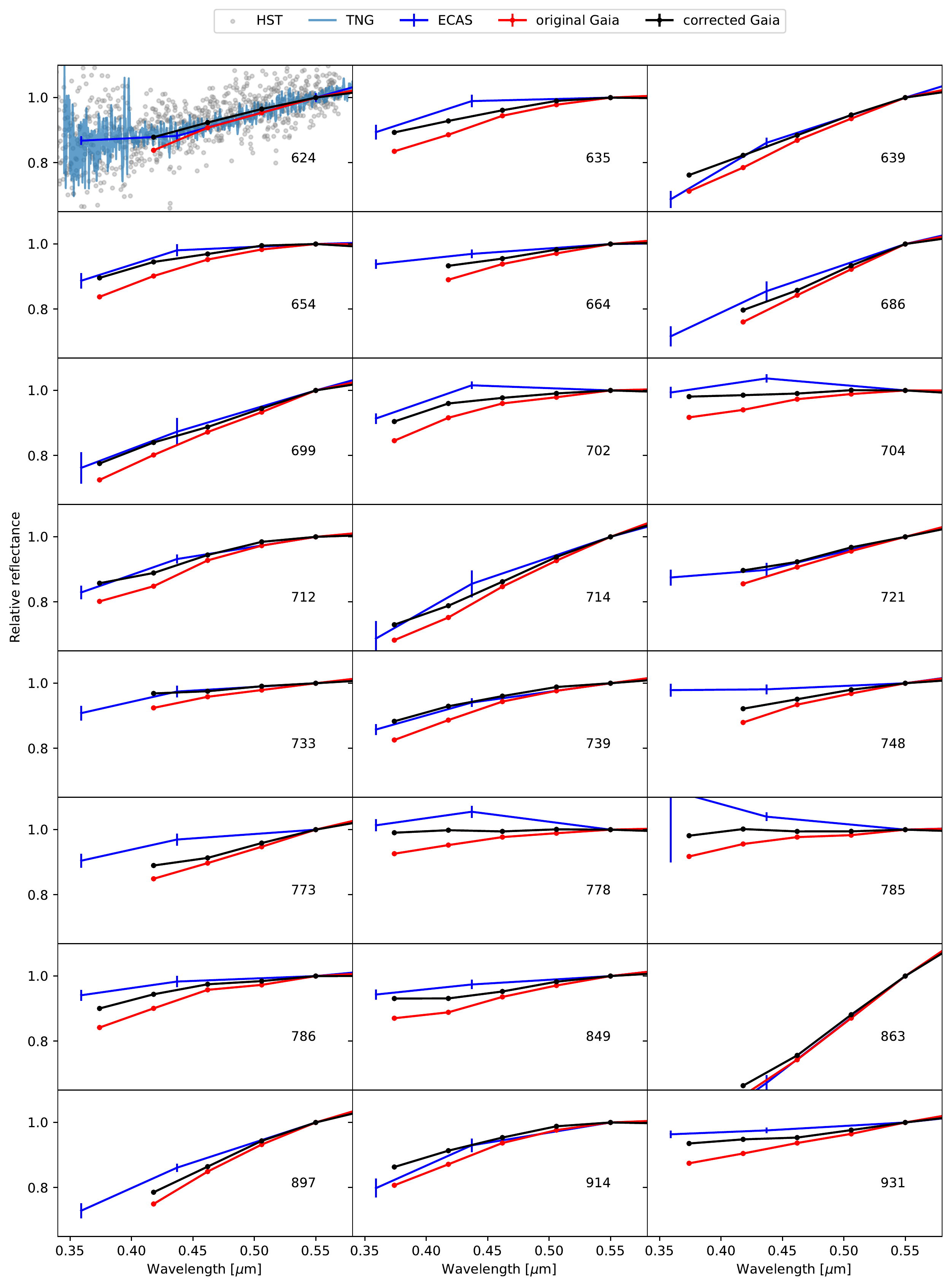}
\end{center}

\clearpage

\begin{center}
\centering
\includegraphics[width=0.95\textwidth]{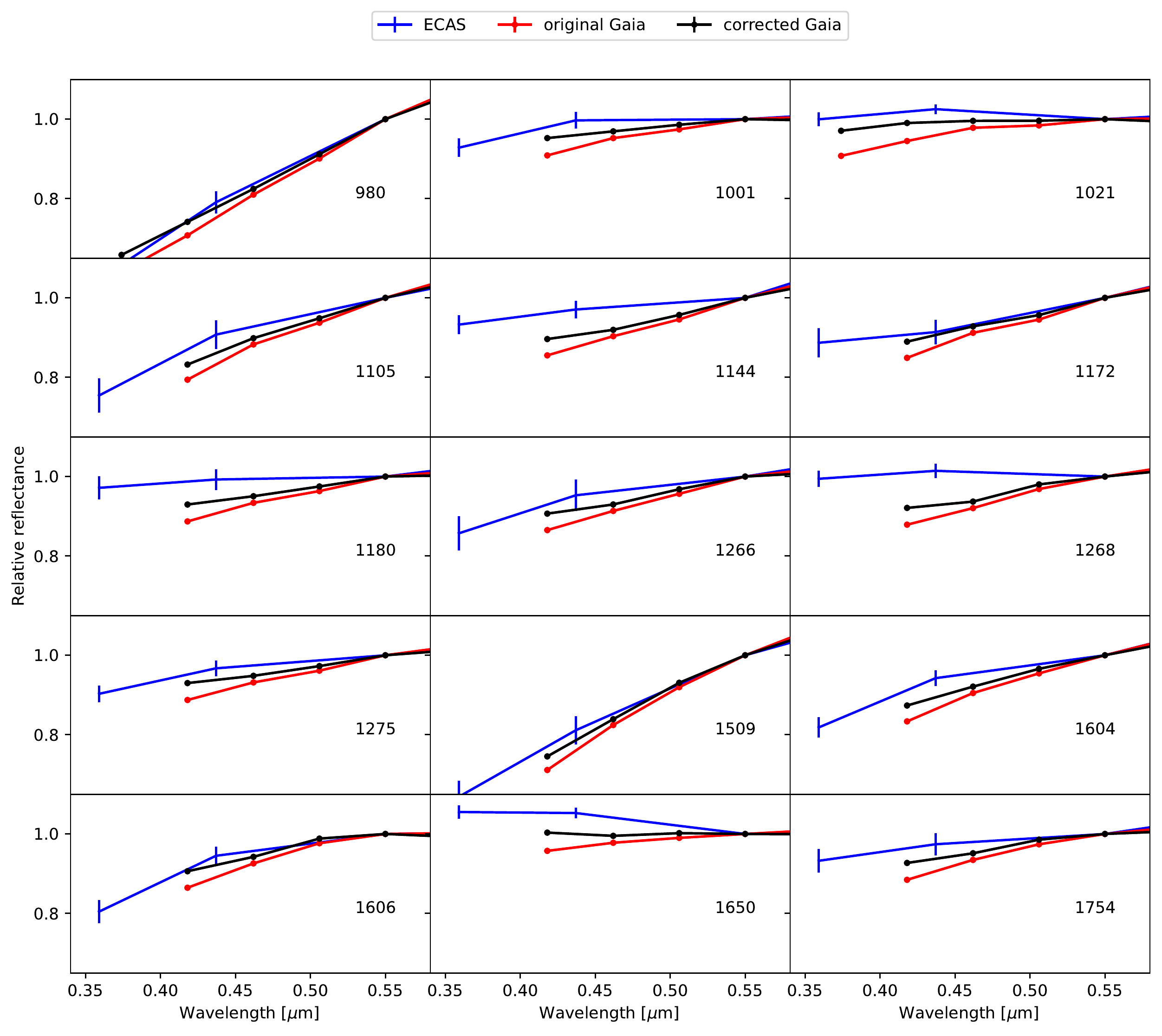}
\end{center}

\end{document}